# Titanium trisulfide (TiS$_3$): a 2D semiconductor with quasi-1D optical and electronic properties

*Joshua O. Island[1]\*, Robert Biele[2], Mariam Barawi[3, †], José M. Clamagirand[3], José R. Ares[3], Carlos Sánchez[3,4], Herre S.J. van der Zant[1], Isabel J. Ferrer[3,4], Roberto D'Agosta[2,5]\*, and Andres Castellanos-Gomez[1,6]\**

[1]Kavli Institute of Nanoscience, Delft University of Technology, Lorentzweg 1, 2628 CJ Delft, The Netherlands.

[2] Nano-Bio Spectroscopy Group and European Theoretical Spectroscopy Facility (ETSF), Universidad del Pais Vasco, 20018 San Sebastian, Spain

[3]Materials of Interest in Renewable Energies Group (MIRE Group), Dpto. de Física de Materiales, Universidad Autónoma de Madrid, 28049 Madrid, Spain

[4]Inst. Nicolas Cabrera, Univ. Autónoma de Madrid, 28049 Madrid, Spain

[5] IKERBASQUE, Basque Foundation for Science, 48013, Bilbao, Spain

[6]Instituto Madrileño de Estudios Avanzados en Nanociencia (IMDEA Nanociencia), Campus de Cantoblanco, 28049 Madrid, Spain.

**Abstract**

We present characterizations of few-layer titanium trisulfide ($TiS_3$) flakes which, due to their reduced in-plane structural symmetry, display strong anisotropy in their electrical and optical properties. Exfoliated few-layer flakes show marked anisotropy of their in-plane mobilities reaching ratios as high as 7.6 at low temperatures. Based on the preferential growth axis of $TiS_3$ nanoribbons, we develop a simple method to identify the in-plane crystalline axes of exfoliated few-layer flakes through angle resolved polarization Raman spectroscopy. Optical transmission measurements show that $TiS_3$ flakes display strong linear dichroism with a magnitude (transmission ratios up to 30) much greater than that observed for other anisotropic two-dimensional (2D) materials. Finally, we calculate the absorption and transmittance spectra of $TiS_3$ in the random-phase-approximation (RPA) and find that the calculations are in good agreement with the observed experimental optical transmittance.

**Introduction**

The isolation of graphene and similar atomically-thin, van der Waals materials has sparked a strong research focus on this broad family which can be exfoliated from bulk layered crystals.[1-3] Despite this growing interest, the research focus has been mainly limited to graphene, boron nitride, and the Mo- and W- based transition metal dichalcogenides.[4-7] These materials, although very different, have largely isotropic in-plane optical and electrical properties. Reduced in-plane symmetry could lead to interesting anisotropic properties furthering the functionalities and applications of two-dimensional (2D) materials. In particular, anisotropic 2D materials would be appealing to fabricate passive optical polarizers and high mobility transistors that benefit from reduced backscattering from hot electrons.[8]

Recently, the layered materials black phosphorus (BP) and rhenium disulfide (ReS$_2$) have shown promising in-plane anisotropy of their electrical and optical properties.[9-11] Harnessing the anisotropic properties of these materials has resulted in applications such as an integrated digital inverter[12], and a linear-dichroic photodetector.[13] Exploration of other layered materials with stronger structural in-plane anisotropy would allow the opportunity to couple the advantages of 2D materials (flexibility, transparency, and large surface to volume ratio) with fully quasi-one dimensional (1D) properties. The transition metal trichalcogenides (with general formula MX$_3$) are a prospective family of materials to achieve this goal due to their reduced in-plane bonding symmetry.[14] Figure 1 shows the crystal structure of a single layer of titanium trisulfide (TiS$_3$), the material under consideration in this study. Differences in the Ti-S bond length along the a (2.65 Å) and b (2.45 Å) axes lead to highly conducting 1D chains along the b-axis and results in strong anisotropic properties.[15,16] The red dashed lines mark the titanium, 1D chains in Figure 1 which are also covalently bonded along the a-axis forming sheets that interact by van der Waals forces. Recent ab initio calculations predict an upper limit for b-axis electron mobilities of more than 10,000 cm$^2$/Vs (higher than MoS$_2$[17]) and a-axis mobilities of more than an order of magnitude less.[15] To date, TiS$_3$ has been exfoliated down to a single-layer (0.7 nm thick) and field-effect transistors and photodetectors have been demonstrated[18, 19-21] but a handling of the in-plane anisotropic properties of 2D, few-layer flakes is still lacking.

Here, specifically, we study the in-plane anisotropy of the electrical and optical properties of few-layer TiS$_3$ flakes. Electrical transport, angle resolved polarization Raman spectroscopy, and optical transmission measurements are used to probe the anisotropic behavior of thin TiS$_3$ specimens. We find that the electrical conductivity shows a marked anisotropy of $G_{max}/G_{min}$ = 2.1 at room temperature and can reach $G_{max}/G_{min}$ = 4.4 at low temperatures. We also note that the Raman spectra of thin flakes show strong in-plane anisotropy that can be used to identify the crystalline orientation of 2D, few-layer TiS$_3$ samples. Interestingly, TiS$_3$ displays strong linear dichroism, optical absorption that depends on the relative orientation between the materials lattice and incident linearly polarized light, with a magnitude much greater than that observed for BP and other anisotropic 2D materials. Finally, to better understand these findings we calculate the absorption and transmittance spectra in the random-phase-approximation (RPA)[22-24] and find that the calculations are in good agreement with our experimental findings.

**Results and Discussion**

TiS$_3$ samples are prepared by mechanical exfoliation of bulk TiS$_3$ material which is synthesized by sulphuration of titanium disks (as reported in Ref. [25]). Recently, we have shown that by varying the temperature of the growth process, control over the morphology of the TiS$_3$ bulk material can be achieved.[19] At a growth temperature of 400 C° the material grows in a sheet-like morphology permitting the isolation of few-layer, 2D flakes while at 500 C° it grows adopting a ribbon-like morphology, characteristic of the MX$_3$ chalcogenides (M = Ti, Zr, Hf and X = S, Se, Te), with high aspect ratio. Figure 2(a) shows an optical image of a 6.4 nm (~9 layers) thick TiS$_3$ nanosheet prepared by mechanical exfoliation of the material grown at 400 C°. In order to study the in-plane anisotropic electrical properties of the exfoliated flakes, 12 electrodes are created (using standard e-beam lithography, Ti/Au evaporation, and lift-off) to measure the electrical properties along different directions in rotational steps of 30 degrees. Figure 2(b) shows an AFM scan of the final device with 12 patterned electrodes.

The reduced thickness of the isolated TiS$_3$ nanosheet allows us to study the electrical transport in a field-effect geometry, using the underlying 285 nm SiO$_2$ as a gate dielectric and the heavily doped silicon as a gate electrode. We modulate the density of the charge carriers by applying a back gate voltage which makes it possible to estimate the field effect carrier mobility. Figure 2(c) shows transistor transfer curves (source-drain current acquired at fixed bias, 100 mV, while sweeping the back gate voltage) measured between two opposing electrodes. We set 0° (dashed white line in Figure 2(b)) as the high conductance axis and plot the transfer curves for the next four pairs of electrodes in steps of 30° in a counterclockwise direction (0° to 120°). Along the b-axis, we estimate an electron mobility of 25 cm$^2$/Vs from the measured transconductance using the standard linear regime, FET mobility calculation (unadjusted for contact resistance).[26] Figure 2(d) shows a polar plot with the angular dependence of the conductance at gate voltages of -40 V, 0 V, and 40 V. The conductances for angles 0° through 120° correspond to the transfer curves in Figure 2(c). The minimum (maximum) conductance directions are determined as the a-axis (b-axis) crystalline directions of the TiS$_3$ nanosheets. These directions correspond to a conductance anisotropy of $G_{max}/G_{min} = 2.1$ as well as an anisotropy in the calculated mobilities of $\mu_b/\mu_a = 2.3$ (see Supporting Information, Figure S1 for a polar plot). The anisotropy in the conductance and mobility represents a more than 30% increase over reported anisotropy values for BP FET devices.[10,27,28] Furthermore, BP shows little change in its anisotropy with decreased temperature whereas we find that the anisotropy in TiS$_3$ is strongly temperature dependent.[10] In Figure 2(e) and 2(f) we plot the transfer curves and angle dependent conductances for the same device measured at a temperature of 25 K. The anisotropy in the mobility increases to $\mu_b/\mu_a$ = (23 cm$^2$/Vs) / (3 cm$^2$/Vs) = 7.6 (see supplement figure S1 for polar plots) and we calculate an increase of the anisotropy in conductance to $G_{max}/G_{min} = 4.4$. Anisotropy values for our few-layer flake devices are slightly lower than the corresponding bulk values.[29,30] We speculate that current spreading and increased Coulomb interactions are the main causes which lead to a decrease in the anisotropy when compared with bulk values. Full temperature dependent measurements of nanosheets of thicknesses from bulk to few layers would help ascertain these discrepancies.

We further characterize the anisotropy of $TiS_3$ by employing angle-resolved polarization Raman spectroscopy, as it has proven to be a very powerful tool to delineate the structural and vibrational properties of 2D materials.[31-33] Figure 3 shows Raman spectra acquired on a $TiS_3$ ribbon grown at 500 °C under different polarization conditions. Note that we first select a nanoribbon to directly identify the crystalline b-axis which is the preferential growth axis. Figure 3(a) shows 5 prominent peaks in the Raman spectra of an isolated nanoribbon sample exfoliated on to a $Si/SiO_2$ substrate. While the peak at ~520 $cm^{-1}$ is due to a Raman mode of the underlying silicon substrate, the other 4 peaks, (occurring at 177 $cm^{-1}$, 300 $cm^{-1}$, 371 $cm^{-1}$, and 559 $cm^{-1}$) correspond to $A_g$-type Raman modes of the $TiS_3$ crystal and are in good agreement with the modes reported for bulk $TiS_3$.[34,35] While all the modes change in intensity with polarization angle (see Supporting Information for angle dependence of all modes), we find that the peak around 370 $cm^{-1}$ is the most sensitive to the relative orientation between the b-axis and the polarization of the excitation laser (top panel of Figure 3(a)). In fact, its intensity is strongly reduced when the polarization of the excitation is parallel to the b-axis. This is evident after rotating the substrate by ~90° and taking a second Raman spectra measurement with the sample now nearly parallel to the excitation/detection polarization (see bottom panel of Figure 3(a)). This finding allows us to determine the crystalline orientation of exfoliated $TiS_3$ nanosheets whose b-axis direction cannot be distinguished at a glance from the material morphology like in the case of the nanoribbons.

In Figure 3(b), the intensity of the 370 cm$^{-1}$ Raman peak (measured on a 3 nm thick, about 4 layer thick TiS$_3$ nanosheet) is displayed at different excitation polarization angles while keeping the detection polarization parallel to the horizontal axis (see Supporting Information for a direct comparison between Raman spectra for the nanoribbon in Figure 3(a) and the nanosheet in Figure 3(b) showing the same Ag-type modes). The minimum of the Raman peak intensity is reached when the excitations polarization forms an angle of 130° with the horizontal axis, indicating the direction of the b-axis in the thin TiS$_3$ nanosheet. Below the polar plot, Figure 3(b) also displays an optical image of the studied flake and an AFM image shown with the same orientation as the optical image. A zoom-in of the AFM image shows how the determined b-axis direction is parallel to the straight edges of the TiS$_3$ nanosheet. In Figure S2, the reader will find polar plots, similar to Figure 3(b), for a TiS$_3$ nanoribbon and a TiS$_3$ nanosheet acquired at different sample orientation angles showing that the minimum occurs when the polarization of the excitation is aligned with the b-axis.

The anisotropy in the optical properties of TiS$_3$ is further characterized using transmission mode optical microscopy. A linear polarizer is placed between the microscope light source and the condenser lens. Transmission mode images are acquired while the polarizer is rotated in steps of ~3°. The transmission is then calculated by normalizing the intensity measured on the TiS$_3$ by the intensity measured on the nearby bare substrate. Figure 4(a) shows a polar plot with the angular dependence of the transmission of a TiS$_3$ nanosheet. Besides the angle values, Figure 4(a) also displays the acquired optical images, highlighting the orientation of the excitation polarization with respect to the nanosheet. The polar-plot clearly shows a marked linear dichroism i.e., a variation in the optical absorption for different polarization angles, for the TiS$_3$ nanosheet. The transmission reaches a minimum value when the excitation light is polarized along the elongated side of the flake, which corresponds to the b-axis. These absorption characteristics are analogous to that of a wire grid polarizer. The high conducting 1D chains of the TiS$_3$ absorb light with a polarization that is parallel to the chain axis (b-axis). The reader can find similar measurements made on a nanoribbon, where the b-axis can be easily determined in the Supporting Information (see Figure S3). The ratio between the b-axis and the a-axis transmission can reach values as high as 30 and decreases for thinner flakes (see Supporting Information, Figure S4). For a direct comparison, we measure the transmission of multilayer BP and MoS$_2$ flakes of similar absolute transmission which present transmission ratios of ≈ 1.4 and ≈ 1, respectively (see Figure S5 in the Supporting Information).

To better understand and reaffirm the linear dichroic behavior found in the optical transmission measurements, we perform density functional theory (DFT) calculations in combination with many-body techniques (see the Methods section) to calculate the absorption spectrum for bulk TiS$_3$. Figure 4(b) shows the calculated absorption spectra for bulk TiS$_3$ when the electric field is aligned parallel to the a-axis (black solid curve) and b-axis (black dashed curve). Across the visible wavelengths, the absorption is much larger when the excitation field is parallel to the b-axis than when it is parallel to the a-axis. The inset in Figure 4(b) shows the calculated transmittance in the a-b plane for red (1.9 eV), green (2.4 eV), and blue (2.72 eV) excitations, agreeing well with the experimental transmittance of the red, green, and blue channels in Figure 4(a).

**Conclusions**

In summary, we present electrical and optical measurements of the in-plane anisotropy of a recently isolated 2D material, TiS$_3$, with 1D-like properties. From electrical measurements we calculate an anisotropy in the in-plane conductivity of 2.1 at room temperature and 4.4 at a temperature of 25 K. Through Raman characterization, we find that the Raman mode at 370 cm$^{-1}$ is sensitive to the orientation of the b-axis relative to that of the excitation polarization. This allows for simple identification of the high conductance b-axis in TiS$_3$ nanosheets. Furthermore, through optical transmission measurements and DFT calculations we show that TiS$_3$ exhibits strong linear dichroism where the ratio in the transmission between the b-axis and a-axis reaches values as high as 30. The strong anisotropic properties of TiS$_3$ set it apart from the commonly studied 2D materials which have largely isotropic properties and makes it an interesting material for future applications.

**Methods section**

Synthesis of TiS$_3$ bulk material: TiS$_3$ has been synthetized by sulfuration of Ti discs (Goodfellow 99.9%, Ø=10mm) which had been previously etched in acid mixture (HF:HNO$_3$, 4:30 wt%) to clean the surface of any oxide impurities. Sulfuration took place in a vacuum sealed ampoule with sulfur powder (Merk, 99.75%) enough to get the sulfur vapor pressure at 500 ºC (~2 bars) and 400 ºC (~0.5 bars) in order to get nanoribbons and nanosheets, respectively. Time of sulfuration was 20 hours. Method was described in Ref.[25].

Fabrication and measurement of FET devices: Few-layer TiS$_3$ nanosheets are isolated and exfoliated onto Si/SiO$_2$ (285 nm) substrates using a viscoelastic stamp (PDMS). Individual nanosheets are selected by optical color contrast and Ti/Au electrodes are patterned using electron beam lithography, thin-film evaporation, and lift-off in warm acetone. Electrical measurements are performed in a Lakeshore probe station in vacuum.

Raman measurements: Raman measurements are performed in a Renishaw system with a 514 nm laser. All spectra are recorded at low powers to prevent laser-induced modification of the samples.

Optical Transmission measurements: A *Nikon Eclipse Ci* optical microscope equipped with a *Canon EOS 1200D* digital camera has been employed to perform the optical transmission measurements. A rotation stage with a linear polarizer has been placed between the illumination source and the microscope condenser to control the polarization of the incident light. In order to determine the polarization angle dependence of the transmission, the polarizer has been rotated in 3º steps and a transmission mode image has been acquired at each step. The transmission is then calculated by dividing the intensity transmitted through the bare substrate and that transmitted by the $TiS_3$.

DFT calculations: Both for Ti and S, the exchange-correlation potential is described self-consistently within the generalized gradient approximation throughout the Perdew–Burke–Ernzerhof's functional (PBE). For S a norm-conserving Martins–Troulliers' pseudopotential is used, while for Ti a norm-conserving Goedecker–Hartwigsen–Hutter–Teter's pseudopotential, including semi-core states for the valence electrons, is used[36,37]. We have relaxed the atomic positions and the unit cell vectors with a residual force after relaxation of 0.001 a.u. by starting from the atomic positions provided in Ref. 17. The kinetic energy cutoff for the plane wave basis set is 180 Ry. The sampling of the Brillouin zone is 10 × 10 × 10 according to the Monkhorst−Pack scheme. First-principles electronic structure calculations and structure optimisation have been performed with the PWSCF code of the Quantum-ESPRESSO package[38]. As DFT tends to underestimate the electronic bandgap, we performed non-self-consistent $G_0W_0$[39,40] calculations in order to get accurate values for the electronic band structure. The local field effects in the screening calculations have been taken into account and we carefully converged the electronic quasi-particle gap. The $G_0W_0$ corrected DFT bands have been used to calculate the absorption spectra in the random-phase approximation (RPA). To construct the kernel in the RPA we have considered 30 valence and 120 conduction bands. The size of the response function has been carefully converged, hence local field effects corresponding to charge oscillations are accurately included. The plane-wave code Yambo[41] is used to calculate the quasi-particle corrections and the optical properties in the RPA.

**Supporting Information**. Room temperature and low temperature field effect mobility polar plots, Raman polarization dependence and direct comparison between nanoribbons and nanosheets, polar plots of the I3 Raman peak with sample rotation, polar plots of the transmittance with sample rotation, thickness dependence of the angular dependent transmittance, comparison of the angular dependent transmittance for $TiS_3$, BP, and $MoS_2$.

**Corresponding Authors**

j.o.island@tudelft.nl

roberto.dagosta@ehu.es

andres.castellanos@imdea.org

**Present Addresses**

† Center for Biomolecular Nanotechnologies, Fondazione Istituto Italiano di Tecnologia, 73010 Arnesano (LE), Italy

**Financial Interests**

The authors declare no competing financial interests.

**Author Contributions**

A.C.G conceived the project. J.O.I and A.C.G. performed the experiments. R.B. and R.D. performed the calculations. M.B., J.M.C., J.R.A., C.S., and I.J.F. synthesized the bulk material. J.O.I., R.B., R.D., H.S.J.Z. I.J.F., and A.C.G wrote the manuscript. All authors contributed though scientific discussion.

**Acknowledgments**

This work was supported by the Dutch Organization for Fundamental Research on Matter (NWO/OCW). R. D'A. and R. B. acknowledge financial support by the DYN-XC-TRANS (Grant No. FIS2013-43130-P), and NanoTHERM (Grant No. CSD2010-00044) of the Ministerio de Economia y Competitividad (MINECO), and Grupo Consolidado UPV/EHU del Gobierno Vasco (Grant No. IT578-13). R. B. acknowledges the financial support of the Ministerio de Educacion, Cultura y Deporte (Grant No. FPU12/01576). AC-G acknowledges financial support from the BBVA Foundation through the fellowship "*I Convocatoria de Ayudas Fundacion BBVA a Investigadores, Innovadores y Creadores Culturales*", from the MINECO (Ramón y Cajal 2014 program, RYC-2014-01406) and from the MICINN (MAT2014-58399-JIN). MIRE Group thanks MINECO (MAT2011-22780) for financial support.

**References**

1   Novoselov, K. *et al.* Two-dimensional atomic crystals. *Proc. Natl. Acad. Sci. U. S. A.* **102**, 10451-10453 (2005).
2   Butler, S. Z. *et al.* Progress, challenges, and opportunities in two-dimensional materials beyond graphene. *ACS Nano* **7**, 2898-2926 (2013).

3    Mak, K. F., Lee, C., Hone, J., Shan, J. & Heinz, T. F. Atomically Thin MoS_ {2}: A New Direct-Gap Semiconductor. *Phys. Rev. Lett.* **105**, 136805 (2010).
4    Schwierz, F. Graphene transistors. *Nat. Nanotechnol.* **5**, 487-496 (2010).
5    Radisavljevic, B., Radenovic, A., Brivio, J., Giacometti, V. & Kis, A. Single-layer MoS2 transistors. *Nat. Nanotechnol.* **6**, 147-150 (2011).
6    Dean, C. *et al.* Boron nitride substrates for high-quality graphene electronics. *Nat. Nanotechnol.* **5**, 722-726 (2010).
7    Georgiou, T. *et al.* Vertical field-effect transistor based on graphene-WS2 heterostructures for flexible and transparent electronics. *Nat. Nanotechnol.* **8**, 100-103 (2013).
8    Abudukelimu, A. *et al.* in *Solid-State and Integrated Circuit Technology (ICSICT), 2010 10th IEEE International Conference on.*   1247-1249 (IEEE).
9    Ho, C., Huang, Y., Tiong, K. & Liao, P. In-plane anisotropy of the optical and electrical properties of layered ReS2 crystals. *J. Phys.: Condens. Matter* **11**, 5367 (1999).
10   Xia, F., Wang, H. & Jia, Y. Rediscovering black phosphorus as an anisotropic layered material for optoelectronics and electronics. *Nat. Commun.* **5**, 4458 (2014).
11   Chenet, D. *et al.* In-Plane Anisotropy in Mono-and Few-Layer ReS2 Probed by Raman Spectroscopy and Scanning Transmission Electron Microscopy. *Nano Lett.* **15**, 5667-5672 (2015).
12   Liu, E. *et al.* Integrated digital inverters based on two-dimensional anisotropic ReS2 field-effect transistors. *Nat. Commun.* **6** (2015).
13   Yuan, H. *et al.* Polarization-sensitive broadband photodetector using a black phosphorus vertical p–n junction. *Nat. Nanotechnol.* **10**, 707-713 (2015).
14   Meerschaut, A. & Rouxel, J. in *Crystal Chemistry and Properties of Materials with Quasi-One-Dimensional Structures*    205-279 (Springer, 1986).
15   Dai, J. & Zeng, X. C. Titanium Trisulfide Monolayer: Theoretical Prediction of a New Direct-Gap Semiconductor with High and Anisotropic Carrier Mobility. *Angew. Chem.* **54**, 7572-7576 (2015).
16   Jin, Y., Li, X. & Yang, J. Single layer of MX3 (M= Ti, Zr; X= S, Se, Te): a new platform for nano-electronics and optics. *Phys. Chem. Chem. Phys.* **17**, 18665-18669 (2015).
17   Cai, Y., Zhang, G. & Zhang, Y.-W. Polarity-reversed robust carrier mobility in monolayer MoS2 nanoribbons. *J. Am. Chem. Soc.* **136**, 6269-6275 (2014).
18   Island, J. O. *et al.* Ultrahigh Photoresponse of Few-Layer TiS3 Nanoribbon Transistors. *Adv. Opt. Mater.* **2**, 641-645 (2014).
19   Island, J. O. *et al.* TiS3 transistors with tailored morphology and electrical properties. *Adv. Mater.* **27**, 2595-2601 (2015).
20   Lipatov, A. *et al.* Few-layered titanium trisulfide (TiS 3) field-effect transistors. *Nanoscale* **7**, 12291-12296 (2015).
21   Molina-Mendoza, A. J. *et al.* Electronic Bandgap and Exciton Binding Energy of Layered Semiconductor TiS3. *Adv. Electron. Mater.* **1**, 1500126 (2015).
22   Nozieres, P. & Pines, D. Electron interaction in solids. General formulation. *Phys. Rev.* **109**, 741 (1958).
23   Ehrenreich, H. & Cohen, M. H. Self-consistent field approach to the many-electron problem. *Phys. Rev.* **115**, 786 (1959).
24   Nozières, P. & Pines, D. Electron interaction in solids. Collective approach to the dielectric constant. *Phys. Rev.* **109**, 762 (1958).

25  Ferrer, I. J., Maciá, M. D., Carcelén, V., Ares, J. R. & Sánchez, C. On the Photoelectrochemical Properties of $TiS_3$ Films. *Energy Procedia* **22**, 48-52 (2012).
26  Horowitz, G., Hajlaoui, R., Fichou, D. & El Kassmi, A. Gate voltage dependent mobility of oligothiophene field-effect transistors. *J. Appl. Phys.* **85**, 3202-3206 (1999).
27  Liu, H. *et al.* Phosphorene: an unexplored 2D semiconductor with a high hole mobility. *ACS Nano* **8**, 4033-4041 (2014).
28  Lu, W. *et al.* Probing the anisotropic behaviors of black phosphorus by transmission electron microscopy, angular-dependent Raman spectra, and electronic transport measurements. *Appl. Phys. Lett.* **107**, 021906 (2015).
29  Gorlova, I. G., Zybtsev, S. G. e. & Pokrovskii, V. Y. Conductance anisotropy and the power-law current-voltage characteristics along and across the layers of the TiS3 quasi-one-dimensional layered semiconductor. *JETP Lett.* **100**, 256-261 (2014).
30  Gorlova, I. *et al.* Magnetotransport and power-law I–V curves of the layered quasi one-dimensional compound TiS 3. *Physica B* **460**, 11-15 (2015).
31  Ferrari, A. *et al.* Raman spectrum of graphene and graphene layers. *Phys. Rev. Lett.* **97**, 187401 (2006).
32  Ribeiro, H. B. *et al.* Unusual angular dependence of the Raman response in black phosphorus. *ACS Nano* **9**, 4270-4276 (2015).
33  Lee, C. *et al.* Anomalous lattice vibrations of single-and few-layer MoS2. *ACS Nano* **4**, 2695-2700 (2010).
34  Galliardt, D., Nieveen, W. & Kirby, R. Lattice properties of the linear chain compound TiS 3. *Solid State Commun.* **34**, 37-39 (1980).
35  Gard, P., Cruege, F., Sourisseau, C. & Gorochov, O. Single-crystal micro-Raman studies of ZrS3, TiS3 and several Zr1–xTixS3 compounds (0< x⩽ 0.33). *J. Raman Spectrosc.* **17**, 283-288 (1986).
36  Hartwigsen, C., Gœdecker, S. & Hutter, J. Relativistic separable dual-space Gaussian pseudopotentials from H to Rn. *Phys. Rev. B* **58**, 3641 (1998).
37  Goedecker, S., Teter, M. & Hutter, J. Separable dual-space Gaussian pseudopotentials. *Phys. Rev. B* **54**, 1703 (1996).
38  Giannozzi, P. *et al.* QUANTUM ESPRESSO: a modular and open-source software project for quantum simulations of materials. *J. Phys.: Condens. Matter* **21**, 395502 (2009).
39  von der Linden, W. & Horsch, P. Precise quasiparticle energies and Hartree-Fock bands of semiconductors and insulators. *Phys. Rev. B* **37**, 8351 (1988).
40  Engel, G. & Farid, B. Generalized plasmon-pole model and plasmon band structures of crystals. *Phys. Rev. B* **47**, 15931 (1993).
41  Marini, A., Hogan, C., Grüning, M. & Varsano, D. Yambo: an ab initio tool for excited state calculations. *Comput. Phys. Commun.* **180**, 1392-1403 (2009).
42  Momma, K. & Izumi, F. VESTA 3 for three-dimensional visualization of crystal, volumetric and morphology data. *J. Appl. Crystallogr.* **44**, 1272-1276 (2011).

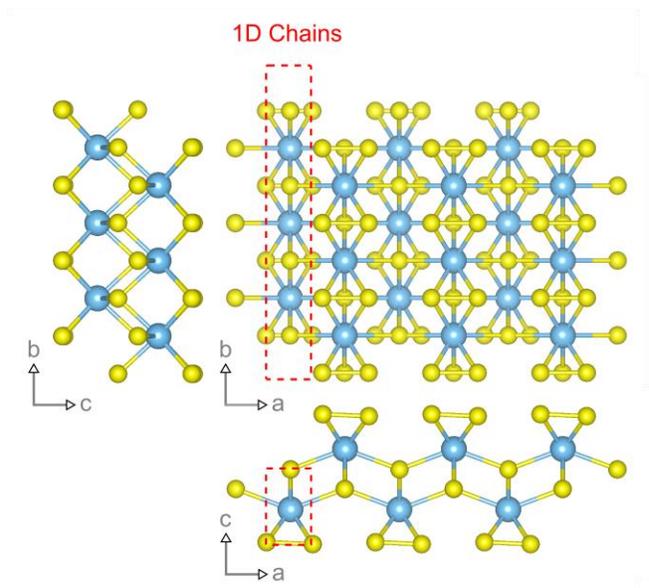

**Figure 1.** Crystal structure of TiS$_3$. The bond lengths between the titanium and sulphur are shorter along the b-axis than along the a-axis. This results in highly conducting chains which lead to strong anisotropic electrical and optical properties. Structure models are produced using VESTA.[42]

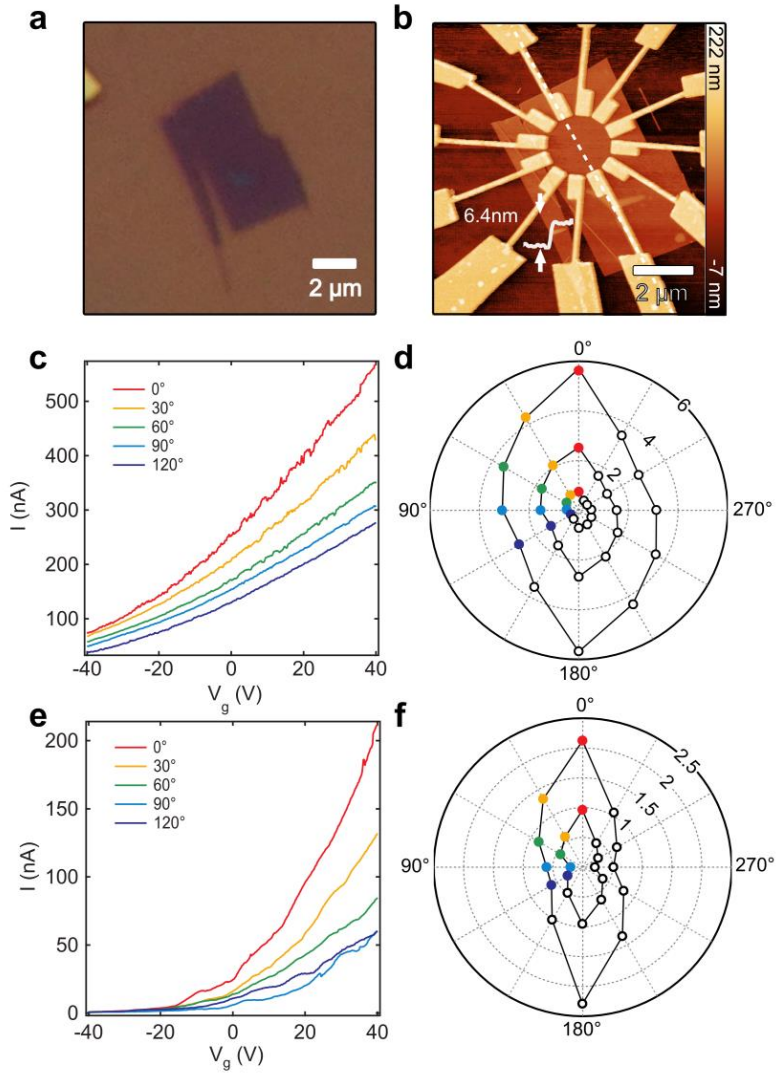

**Figure 2.** (a) Optical image of an exfoliated nanosheet. (b) AFM scan of the same nanosheet after patterning 12 Au/Ti electrodes. (c) Transfer curves measured at room temperature between 5 pairs of electrodes where 0° is designated as the high conductance (b-axis). (d) Polar plot of the room temperature conductance (μS) measured for all 12 pairs of electrodes at back gate voltages of -40 V, 0 V, and 40 V. (e) Transfer curves for the same devices at a temperature of 25 K. (f) Polar plot of the conductance (μS) at a temperature of 25 K and gate voltages of 40 V (outer curve) and 20 V (inner curve).

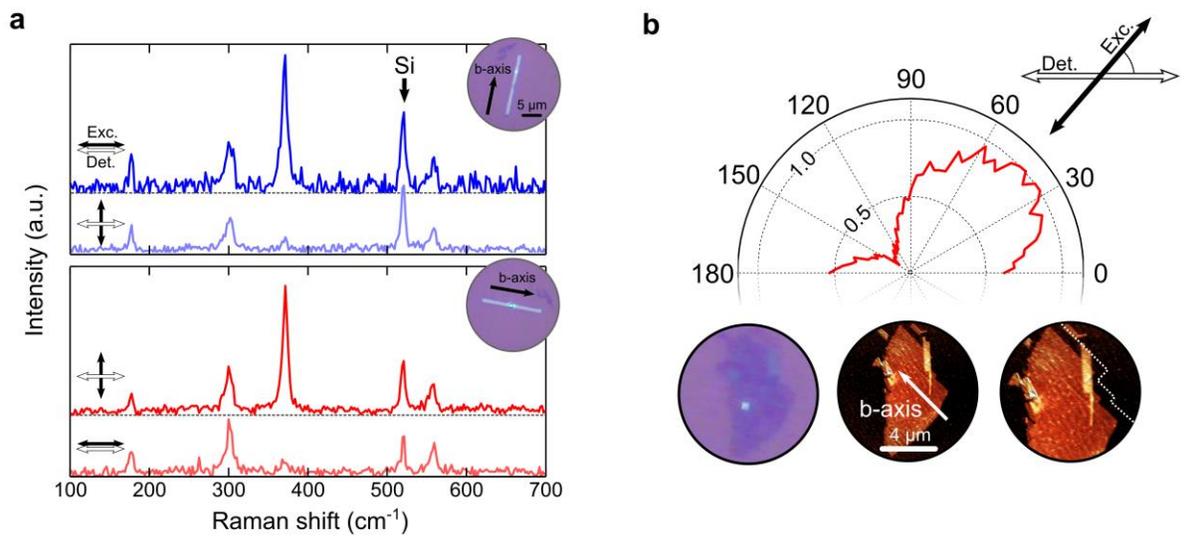

**Figure 3.** (a) Raman spectra of a $TiS_3$ ribbon with horizontal excitation and detection polarization (see the arrows in the insets). In the top (bottom) panel the ribbon has been aligned almost perpendicular (parallel) to the excitation/detection polarization. The insets show the position of the $TiS_3$ ribbon with respect to the illumination polarization. The peak around 370 cm$^{-1}$ shows the most noticeable change with the change of ribbon alignment. (b) Intensity of the 370 cm$^{-1}$ Raman peak of a 3 nm thick $TiS_3$ flake (3-4 layers) as a function of the excitation polarization angle (the detection polarization is fixed along the horizontal axis). The minimum intensity occurs when the excitation polarization is parallel to the b-axis of the flake. (Bottom panels) optical and atomic force microscopy images of the studied flake. The determined b-axis is in good agreement with the straight edges of the $TiS_3$ flake.

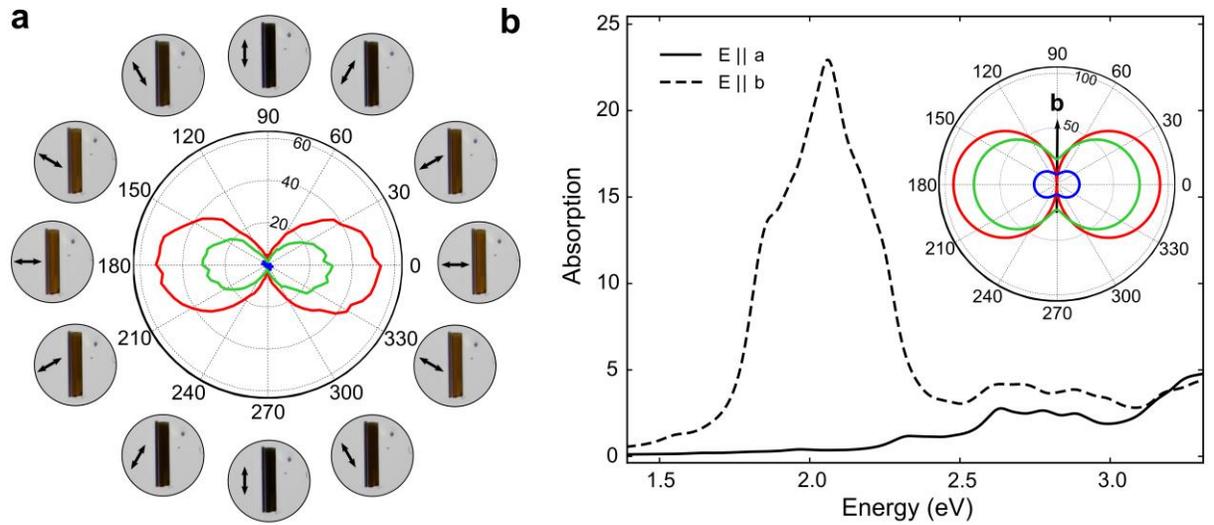

**Figure 4.** (a) Transmittance of the red, green, and blue channels as a function of the excitation polarization angle. (b) Calculated absorption spectra when the field is aligned parallel to the b-axis (green) and a-axis (red). The inset shows the transmittance in the a-b plane for energies red (1.9 eV), green (2.4 eV), and blue (2.72 eV) excitations.

# Supporting Information

Titanium trisulfide (TiS$_3$): a 2D semiconductor with quasi-1D optical and electronic properties

Joshua O. Island*, Robert Biele, Mariam Barawi, José M. Clamagirand, José R. Ares, Carlos Sánchez, Herre S.J. van der Zant, Isabel J. Ferrer, Roberto D'Agosta*, and Andres Castellanos-Gomez*

Supporting Information Contents
1. Room temperature and low temperature field effect mobility
   Figure S1: Polar plot of the field effect mobility.
2. Raman polarization dependence and direct comparison between nanoribbons and nanosheets
   Figure S2: Polar plots and Raman spectra for nanoribbons and nanosheets
3. Polar plots of the I3 Raman peak with sample rotation
   Figure S3: Optical images of exfoliated nanoribbon and nanosheet samples with corresponding polar plots of the normalized I3 Raman peak.
4. Polar plots of the transmittance with sample rotation
   Figure S4: Polar plots with corresponding optical images for four sample rotations.
5. Thickness dependence of the angular dependent transmittance
   Figure S5: Polar plots of the transmittance for different thicknesses.
6. Comparison of the angular dependent transmittance for TiS$_3$, BP, and MoS$_2$
   Figure S6: Polar plots of the transmittance for TiS$_3$, BP, and MoS$_2$.

1. Room temperature and low temperature field effect mobility

The field effect mobility is estimated using the calculated transconductance (dI/dVg) as described in the main text. Figure S1(a) shows a polar plot of the mobility for each pair of electrodes for the device shown in Figure 2(b). The b-axis mobility reaches 25 cm$^2$/Vs and the minimum mobility is at 120° with a value of 11 cm$^2$/Vs. The anisotropy ratio is then 2.3 for the room temperature measurements. Figure S1(b) shows the same polar plot oft he mobility at a measurement temperature of 25 K. The ratio increases to $\mu_b/\mu_a$ = 23 cm$^2$/Vs / 3 cm$^2$/Vs = 7.6.

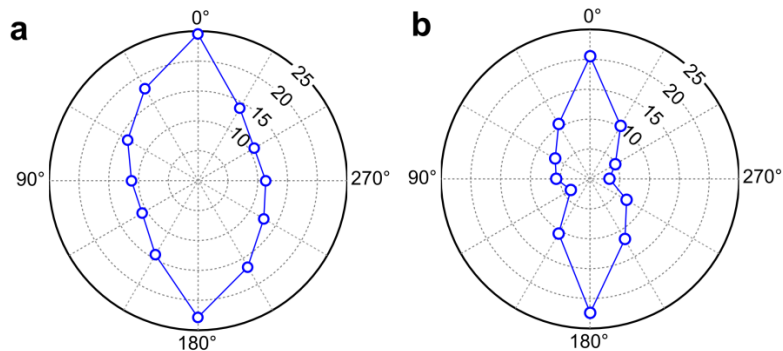

Figure S1: (a) Polar plot of the estimated FET mobility at room temperature. (b) Polar plot of the estimated FET mobility at 25 K.

2. Raman polarization dependence and direct comparison between nanoribbons and nanosheets

In Figure S2(a) we show the angle dependence of all the Raman modes for the nanoribbon and nanosheet in Figure 3 of the main text. While all the modes show some dependence, the peak at 370 cm-1 (blue curve) is most apparent. In Figure S2(b) we directly compare the Raman spectra for the nanoribbon and nanosheet presented in Figure 3 of the main text. This shows that both samples, grown at different temperature, present the same Raman spectra.

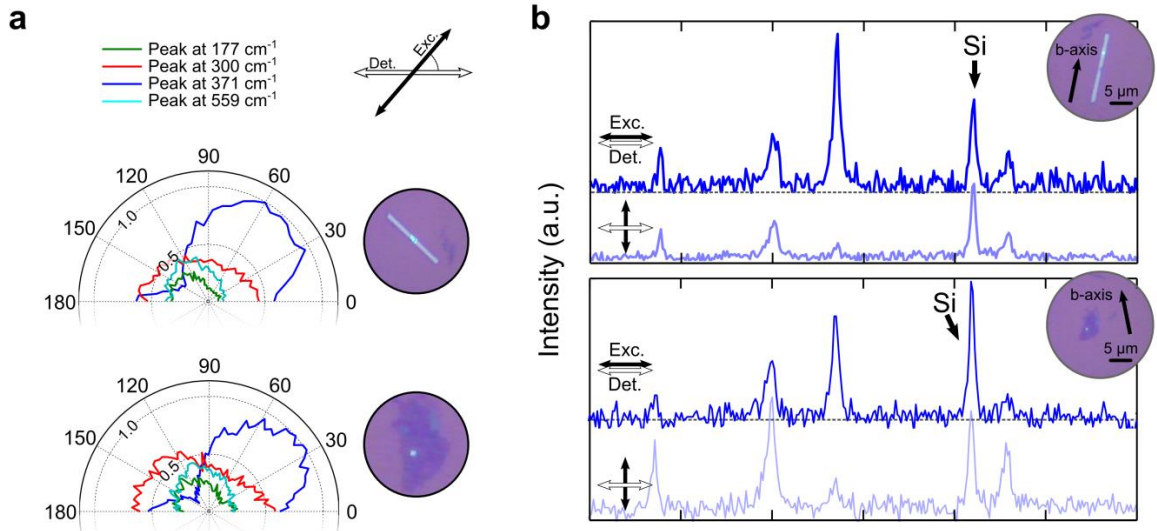

Figure S2: (a) Intensity of all the Raman modes shown in Figure 3 as a function of the excitation polarization angle (the detection polarization is fixed along the horizontal axis). The minimum intensity occurs when the excitation polarization is parallel to the b-axis of the flake. The top panel shows the results for the nanoribbon in Figure 3(a) and the bottom panel shows the results for the nanosheet in Figure 3(b) of the main text. (b) Raman spectra of a $TiS_3$ ribbon with horizontal excitation and detection polarization (see the arrows in the insets) and for a nanosheet (bottom panel). The peak around 370 cm$^{-1}$ shows the most noticeable change with the change of ribbon alignment.

3. Polar plots of the I3 Raman peak with sample rotation

Here we show that there exists a direct correlation between the minimum of the I3 Raman peak and the b-axis of exfoliated $TiS_3$ samples by rotating the sample. First we select a nanoribbon which allows us to clearly identify the b-axis which is the preferential growth direction for $MX_3$ chalcogenides. Figure S3(a) shows optical images of an exfoliated nanoribbon sample and corresponding polar plots of the normalized intensity of the I3 Raman peak (370 cm$^{-1}$) as the angle between the excitation and detection polarizations are changed. It can be seen that the minimum of the I3 peak is achieved when the excitation polarization is parallel with the b-axis of the nanoribbon. With this in mind, the b-axis of exfoliated few-layer $TiS_3$ nanosheet samples (grown at 400 C) can be easily determined. Figure S3(b) shows the same data set for an exfoliated nanosheet. The b-axis for this sample is found to be along the edge of the flake marked with a black line in the optical images.

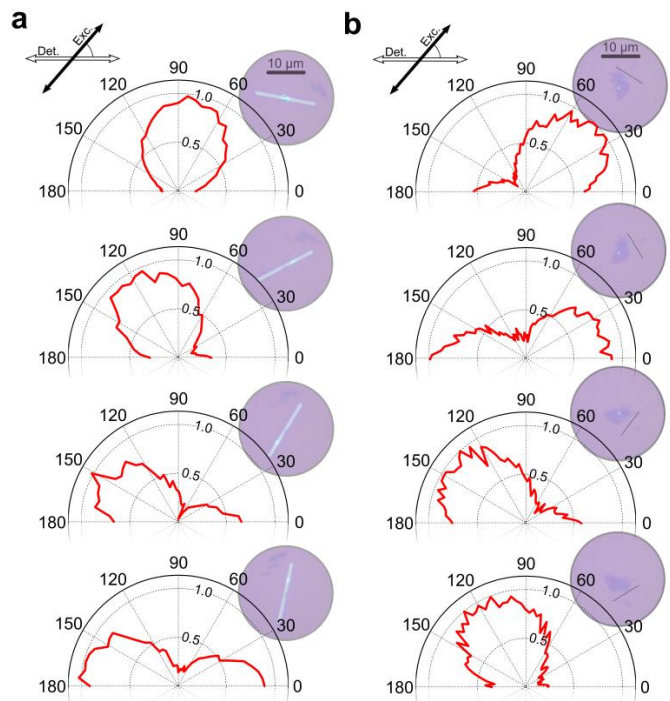

Figure S3: (a) Normalized intensity of the I3 Raman peak of a $TiS_3$ ribbon as a function of the angle between the excitation and detection polarization. The ribbon has been rotated and the measurement has been repeated several times to illustrate that the minimum of the normalized I3 peak is reached when the excitation polarization is parallel to the $TiS_3$ b-axis. (b) Normalized intensity of the I3 Raman peak of a $TiS_3$ nanosheet as a function of the angle between the excitation and detection polarization. The nanosheet has been rotated and the measurement has been repeated several times to illustrate that minimum of the normalized I3 peak is reached when the excitation polarization is parallel to the $TiS_3$ b-axis.

4. Polar plots of the transmittance with sample rotation

Figure S4 shows that the polar dependence of the transmittance follows subsequent rotations of the sample. The minimum of the transmittance is found to correspond with the b-axis. As the sample is rotated, the corresponding polar dependences rotates as well.

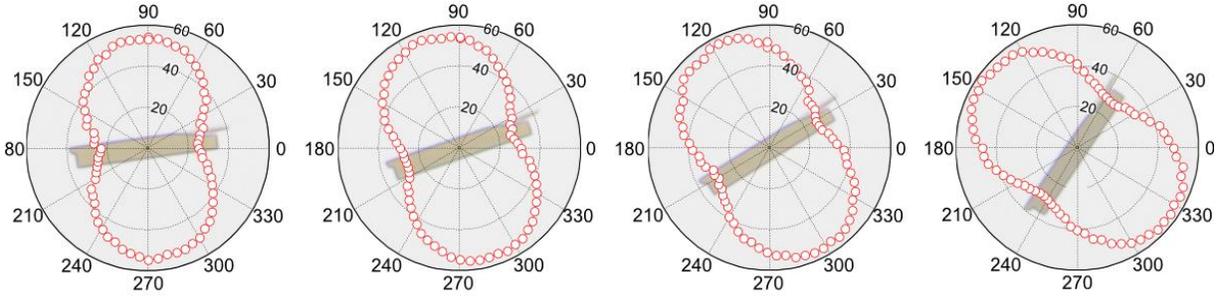

Figure S4. Transmittance as a function of the excitation polarization angle for a TiS$_3$ wide ribbon, rotated at different angles. The angular dependence of the transmittance follows the rotation of the flake.

5. Thickness dependence of the angular dependent transmittance

Figure S5(a-d) shows the angular dependence of the transmittance for decreasing thicknesses of nanosheet samples. The linear dichroism becomes weaker for thinner samples. This can be directly appreciated in Figure S4(e) where we plot the ratio of the maximum and minimum transmittance as a function of the minimum transmittance (thickness). The ratio decreases for thinner samples.

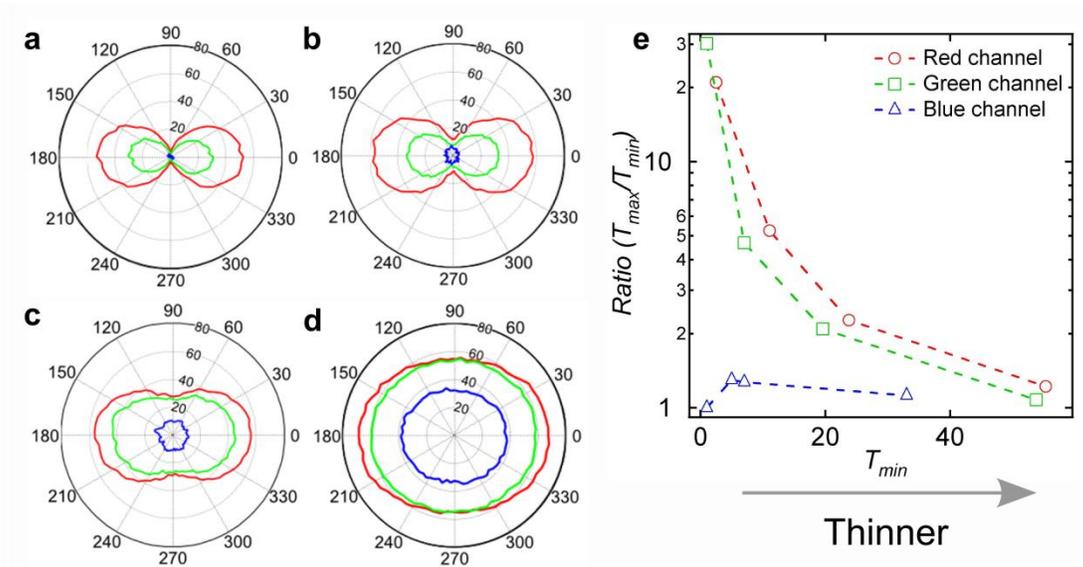

Figure S5. Transmittance measured from the red, green and blue channel of the camera, measured for TiS$_3$ samples with different thicknesses (from thicker to thinner).

6. Comparison of the angular dependent transmittance for TiS$_3$, BP, and MoS$_2$

Figure S6 shows polar plots of the transmittance as a function of excitation angle for samples of TiS$_3$, BP, and MoS$_2$ having comparable overall transmittance. It can be seen that the TiS$_3$ sample has the strongest modulation of the transmittance with a b-axis to a-axis ratio of 30 compared with a ratio of 1.4 for BP. MoS$_2$ shows little modulation as expected.

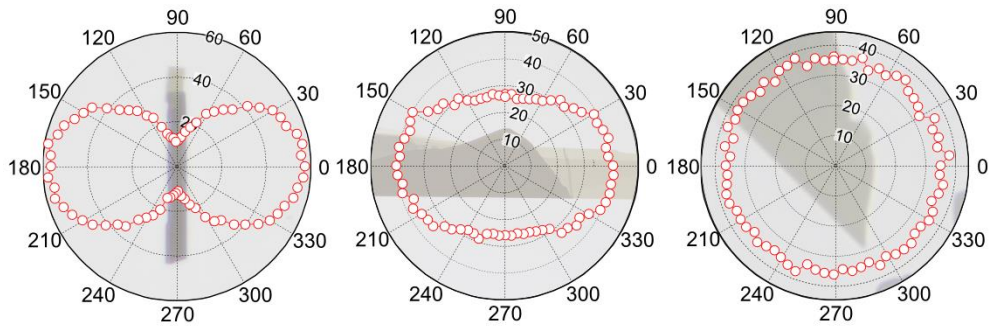

Figure S6. Comparison between the angular dependent transmittance of TiS$_3$ (left), BP (middle) and MoS$_2$ (right).